# TRANSACTION COSTS IN COLLECTIVE WASTE RECOVERY SYSTEMS IN THE EU

**Head Assist. Prof. Shteryo S. Nozharov, PhD[1]**
*UNWE – Sofia,*
*Department of Economics*

**Abstract:** The study aims to identify the institutional flaws of the current EU waste management model by analysing the economic model of 'extended producer responsibility' and collective waste management systems and to create a model for measuring the transaction costs borne by waste recovery organizations. The model was approbated by analysing the Bulgarian collective waste management systems that have been complying with the EU legislation for the last 10 years. The analysis focuses on waste oils because of their economic importance and the limited number of studies and analyses in this field as the predominant body of research to date has mainly addressed packaging waste, mixed household waste or discarded electrical and electronic equipment. The study aims to support the process of establishing a 'circular economy' in the EU, which was initiated in 2015.

At the end of 2015 the European Union made considerable efforts to change its economic model with a proposal for legislative changes aiming to create a 'circular economy' (COM/2015/0595). The grounds for and objectives of the proposal were to increase resource efficiency, to reduce public waste disposal costs, and to improve health and environmental policies.

In 2011 The European Commission published a short list of raw materials (e.g. tungsten, beryllium, germanium, etc.) that are strategic for the European economy (COM/2011/0025). The list shows that the sources of imports of these raw materials are exclusively from non-EU countries. The increased price volatility on the international commodity markets and the frequent political changes in these countries pose a high risk to the European economy. At the same time these elements are present in household waste generated on the territory of the EU. These facts have led to increased calls for policy responses to improve waste management as particularly important for achieving high economic growth based on innovative technologies that require scarce resources.

---

[1] E-mail: nozharov@unwe.bg

The 'extended producer responsibility' principle (Directive 2008/98/EC, Art. 8) is an important economic instrument in the field of waste management. One of the ways of its implementation is through establishment of 'collective waste management systems', which combine the efforts of many producers, dealers, and importers of waste-generating products. These joint efforts aim to reduce the financial burden borne by the companies and at the same time to improve public wealth by reducing the waste flows and improving the efficiency of their management.

The model was adopted in 2008 (Directive 2008/98/EC) and has been in force for already ten years but its effects are controversial. The explanatory memorandum of the circular economy proposal quoted above states that an additional 600 million tons of EU-generated waste could be recycled or reused. This quantity is about 25% of the total waste, which pollutes the environment and could be a source of important secondary raw materials for the economy. The inefficient utilization of waste on the one hand and the depletion of the available natural resource on a global scale are sufficient grounds for scientific research in this field.

The aim of this paper is to identify the reasons for the low efficiency of the European waste management model by means of institutional economy methods and to propose a model for measuring the transaction costs of collective waste management systems.

The study focuses on collective waste recovery systems based on the principle of extended producer responsibility. The scope of the study was reduced to include only the streams of waste oils and other petroleum products because of the great diversity of waste types. Another reason for narrowing the scope is the small number of analyzes specifically related to this waste stream. Most scientific publications focus on packaging waste and mixed household waste. The study is limited to the territory of Bulgaria as an EU Member State, but the conclusions could be used for any of the other European countries.

The results of the study may be used to improve the efficiency of the current waste management system and provide more valuable raw materials for the economy. They may also be used to optimize the economic instruments of the new legislation for transition to a 'circular economy' and avoid the flaws of the current model.

## 1. Literature Review

Despite the little time that has elapsed since the publication of the circular economy proposal of the European Commission in 2015, there are already quite a few scientific publications related to it. One of the latest is by Richard Hughes (2017). It reviews the current regulatory institutional environment in which the circular economy should develop. A key conclusion regarding the new element of this environment is that the regulations of the EU would not be effective without establishing official technical standards for their implementation. These would allow for a strict control and would eliminate the opportunities for different interpretations of the EU regulations by the different Member States as is currently the case. Hughes' publication (2017) does not include a detailed analysis of the principle of extended producer responsibility and does not specifically address collective waste recovery systems and therefore does not coincide with the purpose and methodology of this study.

An analysis of the economic significance of the circular economy model is included in the publication of Ying and Li-jun (2011). The paper provides a detailed comparison between green supply chain management and traditional supply chain management. According to the authors, traditional supply chain management is based on the classic concept of profit maximization as the company's primary goal, which excludes any increase of green management costs. However, they believe that the companies with green supply chain management can have competitive advantages. First of all, rising the green consumption awareness of the market environment (through media, schools, etc.) and explaining its benefits to society will result in higher demand for the products of the companies with green management. The second advantage is seen as establishment of green corporate culture through more rigorous public control over the adherence to environmental regulations and raising corporate employees' awareness of the benefits from a green image of their company. The third advantage is the establishment of strategic cooperation for long-term supply only with green management suppliers and avoidance of traditional management suppliers. The fourth advantage is seen as improvement of the international image of the country and, accordingly, a higher demand for the products of export-oriented companies. The quoted publication of Ying et al. (2011) does not provide a detailed analysis of the principle of extended producer responsibility and does not specifically address collective waste recovery systems and therefore does not duplicate the aim and methodology of this study.

The extended producer responsibility and collective waste recovery systems are addressed in a number of publications but none of them duplicates this study in terms of objectives, methodology, and conclusions. One such paper was published by M. Dubois (2012). It summarizes a number of previous publications on the issue and addresses extended producer responsibility by highlighting some criticisms of weak incentives for waste prevention and recycling. These critical conclusions are related to static goals and the lack of sufficient economic instruments or coordination thereof. According to the author waste recycling or recovery can be stimulated by levying taxes on collective systems for uncollected waste fractions or on producers for the use of primary natural resources in their production process. He also believes that large subsidies for separate collection reduce incentives for ecodesign of products, which limits the potential waste at the stage of product design.

Dubois (2012) discusses collective waste management systems as well. He believes that they provide opportunities for minimization of producers' transaction costs. The author points out that the main disadvantage of these systems is the flat-rate contribution for the recovery of different types of products (e.g. refrigerators) at the end of its life cycle regardless of the different quantities of materials used for their production, which does not stimulate ecodesign.

As far as Dubois (2012) only notes that transaction costs can be considered a waste management instrument, they are viewed in relation to the cost of monitoring of illegal waste disposal. The costs incurred from collusions among between the participants within the collective waste management systems may be defined as social costs.

Transaction costs for waste management are also indirectly referred to by other authors, such as Calcott and Walls (2005). They define transaction costs as one of the many elements of a general equilibrium model of production, consumption, recycling and landfilling. However, they associate these costs only with the effect of taxes and subsidies related to waste management. They believe that in a market-oriented model the transaction costs would be much higher than in the tax and subsidy model. Transaction costs for waste management are considered in a similar way by Shinkuma (2007) and Ino (2011).

## 2. Structural analysis of transaction costs in collective waste recovery systems

This analysis of the structure of transaction costs in collective waste recovery systems is based on the model of transaction costs structure for collective waste recovery systems introduced by Coggan, Whitten and Bennett (2010), which defines the characteristics of the transaction for the environmental good, the nature of the transactors, and the current institutional environment. Although this model is aimed primarily towards environmental policy selection and enforcement by governmental institutions, it can be used as a foundation and modified for the purposes of our analysis.

**Transaction Characteristics**

*General definition of transaction.* According to the Directive on Waste (Directive 2008/98/EC, Art. 21, Para. 3), waste oils 'regeneration' is given priority as a method for their recovery.[2] Regeneration of waste oils is defined in the same Directive (Directive 2008/98/EC, Art. 3, Para. 18) as 'any recycling operation whereby base oils can be produced by refining waste oils.'[3] The resulting 'base oils' cease to be waste (Directive 2008/98/EC, Art. 6, Para. 1) and become raw materials for production of engine oils, gearbox oils, and other industrial oils intended for sale. Therefore, the product (base oils) from regeneration of waste oils is a commodity with economic value and the resulting market price. Thus the object of the analysed transaction is waste in the beginning of the transaction and a commodity at the end of the transaction.

The institutional features of a transaction will be defined in terms of the specific characteristics of the assets, the frequency and the duration of the transaction.

First we shall analyse the *specific characteristics* of the assets.

Using the definition of Williamson (1996), the specific characteristics of transaction assets are defined as relatively high. The recovery of waste oils in Bulgaria through regeneration is carried out only by two enterprises, which have the necessary utilities and permits. This process requires sophisticated utilities and highly qualified staff, which requires large initial investments. It also requires a lot of time for obtaining the various required permits (e.g. IPPC permit under the Environment Protection Law). The regeneration process requires specific inputs (e.g. additives) and activities (Directive 2008/98 / EC, art. 21, para. 3). These specific characteristics of the asset increase the associated transaction costs.

Secondly, we shall analyse transaction *frequency and duration*. The specific characteristics of the asset determine to a large extent the frequency and duration of the transaction. In terms of statutory procedures (Law on Waste

---

[2] The provisions of the Directive were transposed in Bulgaria's legislation with Art. 3 of the Ordinance on Waste Oils and Waste Oil Products /NOMON/. Regeneration priority is imposed in the IPPC permit of each subcontractor. According to Art. 28, Para. 1, item 2 of the Ordinance /NOMON/ when subcontractors are issued IPPC permits according to Chapter Seven, Section II of the Environment Protection Law (EPL) (the predominant case), they are legally obliged to regenerate the waste oils.

[3] These provisions of the Directive were transposed in Bulgaria's legislation with §1, Item 2 and Item 22 of the Ordinance /NOMON/.

Management, 2012, Art. 18), the transaction is carried out once a year and lasts throughout the year. The duration is determined by the large quantities to be collected and processed (40% of the oils and petroleum products sold on the national market). On the one hand, the transaction is recurring, which implies accumulation of experience and establishment of detailed contractual documentation by the transactors. On the other hand, the transaction is associated with large amounts of a specific asset with a long life cycle. Both factors require a complex monitoring process. The transaction is also affected by third parties (the government, NGOs) that control the process or are interested in its monitoring. Their behavior is difficult to predict by the main contractors - the contracting authority and the processor. Therefore, the frequency and duration of the transaction do not have a significant effect on the amount of transaction costs.

**Transactor Characteristics**

*General characteristics* of collective waste recovery systems. Collective waste recovery systems operate under the conditions of publicly available resources, which entails an obligation for these systems to comply with the principle of public efficiency in their activities. Although they are registered as companies under the commercial legislation, the requirements for their registration in Bulgaria are set out in the Law on Waste Management (2012), which substantially alters their character as a classical commercial company.[4] They are prohibited to share profits, to issue bonds and stocks with dividend coupons, and to grant loans and guarantee loans to third parties as well as take promissory liabilities and issue bearer shares.[5] Moreover, they cannot be transformed.[6] According to the provisions of the Law on Waste Management, they cannot carry out activities other than to manage and/or independently carry out the activities of separate collection, recycling and recovery of widespread waste.

In reality, all businesses work for profit and are not subject to restrictions by the Law on Commerce and although the collective recovery systems are registered as business organizations, they are only allowed to carry out certain environmental activities for public benefit. The non-distribution of profits and all other restrictions rank these systems closer to the non-governmental organizations, and their environmental goals, combined with the prohibition to work for profit, identify them as organizations for public benefit. This implies that all activities of the collective recovery systems must comply with the of public effectiveness principle.

---

[4] See Art.16, item 3, Art.17, Para.1 and Para. 2, §1, item 16 of the SP of the Law on Waste Management (2012).

[5] They are prohibited to share profits, to issue bonds and stocks with dividend coupons, and to grant loans and guarantee loans to third parties as well as take promissory liabilities and issue bearer shares.

[6] They cannot be transformed by consolidation, merger, division, separation of the privately owned company or transfer of all assets of the sole shareholder except in cases of consolidation or merger of utilization organizations.

Transactors' specific characteristics are defined in terms of their limited rationality and opportunism.

Transactors' *limited rationality* is related to the complexity of the problems they have resolve and the volume of information they have to analyse (Coggan et al. 2010). The specific characteristics of the asset subject to this specific transaction presume complex problems and large volumes of information the transactors have to deal with. In the context of constant changes to waste regulations at national and European level, transactors' behavior has to be adjusted frequently. Transactors need time to become familiar with and gain experience on the revised regulatory requirements. At the same time the extremely small number of employees (two employees on average) of the Bulgarian collective systems for utilization of waste oils limits the capacity of these organizations to adapt to rapid changes in their business environment when they have to process large volumes of technical and regulatory information and they are more likely to make wrong decisions and this increase their transaction costs.

Transactors' *opportunism* is defined as transactors' possibility to take advantage of false information, withheld information, or flaws in the system of monitoring (Falconer and Saunders, 2002). The specific physical characteristics of the asset hinders control over the processes for collection and regeneration of waste oils and their conversion into base oils as well over any possible follow-up operations with such oils. On the other hand, the limited administrative capacity of the contracting entity (collective system) does not allow for on-site control over the contractor. These characteristics create opportunities for objective misinformation of the contracting authority, and contractors can take opportunistic advantage in order to maximize their profit from the transaction. Moreover, the contracting parties do not share common values. In terms of their organisation, collective utilisation systems are similar to NGOs because the scope of their operations is limited by the environmental legislation and they are not allowed to share profit while the contractors are commercial companies established for profit maximization profits and operating in a duopoly with potential for market domination.

**Uncertainty of the current institutional environment**

EU waste management directives do not specify a model of collective systems that shall be uniformly applied across all member states as each state shall stipulate in its national legislation (European Commission, 2012) how to achieve the goals set by the EU. The general waste oils management rules for are set out in the Waste Framework Directive (Directive 2008/98/EC, Art. 3, Para. 3, 18; Art. 21.)[7] According to the Bulgarian legislation (Law on Waste Management (2012), Art. 14), collective waste utilization systems have two options – either to perform their duties individually, provided that they have the necessary capacity to do so, or to perform their duties through collective systems after signing a contract with utilization organizations possessing the required permits.

None of the Bulgarian collective systems dealing with waste oils has created its own capacity throughout the studied period (2007-2017). According to data from the Commercial Register (2017), the average value of their tangible fixed assets (TFA) is EUR 2000 with the highest value being EUR 8 000, which indicates zero technical capacity. This is why all Bulgarian collective waste oil systems apply the second option by entering into contracts with subcontractors - commercial companies with the respective capacity and permits. These contracts have standard general terms and are concluded in terms of the current market conditions and prices.

A major problem with those standard contracts is the vague definition of their subject matter (the purpose and outcome of the performance.)

Typically, the subject matter of such a contract is defined as: the contracting authority assigns and the contractor undertakes to carry out activities for *collection, transportation, storage and utilisation of waste oils* in accordance with the Ordinance on Waste Oils and Waste Oil Products (NO-MON, 2013). The key issue for clarifying the subject matter of such a contract and the due performance by the subcontractor is the definition of the said '*utilisation of waste-waste oils*'.

The established practice in Bulgaria for the last ten years has been for contractors to issue written statements to the contracting authority (collective system) that hey have fulfilled their contractual obligation to utilize the quantity of waste stipulated for in their contracts. This statement is then submitted by the collective system to the Ministry of Environment and Water. On the grounds of this evidence and an accompanying audit report the ministry exempts the members of the collective system from paying a product fee to the state budget (Waste Management Act 2012, Art. 59, para. 3). Exemption shall be effected by an official order of the Minister, which is published in a special registry of the Ministry of Environment and Water. The order to the Minister of Environment and Water completes the contractual relations between the collective system and the contractor who has issued the statement and the contractor is paid for the rendered services.

---

[7] The Framework Directive (Directive 2008/98/EC) stipulates that waste oils are a sensitive waste group and that member states shall report periodically information on the management of waste oil to the Commission (Art. 28, para. 3 (b); Art. 37).

Therefore, the fulfilment of the contract obligation for recovery of waste oils by the contractor (a company that has the necessary facilities and permits) is proved only on the grounds of the statement issued by the same contractor. The contracting authority (a collective waste recovery system does not have such facilities) submits this statement to the ministry as a proof that it has fulfilled its environmental obligations and has to pay the contractual price to the contractor.

At the same time, the contractor owns the product from the recovery process (base oil) and sells it to third parties as a market commodity. Thus the collective system pays both the raw material and the production costs for the resulting base oil to the contractor who receives the entire profit from the final product at no virtually no cost. This opportunism results from vague institutional rules, generates losses of public welfare, and forms a certain social cost.

### 3. A model for measuring the transaction costs borne by collective waste recovery systems

The model for measuring the transaction costs in the operation of collective waste recovery systems is based on the structural analysis described in the previous section and the model proposed by Collins and Fabozzi (1991). For our purposes their model was modified substantially because it was intended to measure the cost of financial (stock) transactions and its determinants do not meet the requirements of the environmental economics. We have added a second level to the model, which will take into account the social public costs (the impact on public welfare) associated with the measured transaction costs. The new model includes the following variables:

**Social Public Costs (SPC)** = *the quantity of regenerated waste oil with collective systems – the quantity of regenerated waste oil without collective systems.*

**Transaction Costs (TrC)** = *Fixed Costs (FtrC) + Variable Costs (VtrC).*

**FtrC** = *Administrative fixed costs + Market fixed costs.*

**Administrative fixed costs** = *costs associated with the bank guarantee required for the permit + costs associated with the annual performance audits + costs for keeping the required control documentation and assistance during inspections from the public control bodies.*

**Market fixed costs** = *costs associated with the control over contractor's performance + costs associated with the communication among the members of the collective system.*

**VtrC** = *Performance costs + Alternative costs.*

**Performance costs** = *the agreed price for the services rendered by contractors – the price if the collective system has its own recovery facilities.*

**Alternative costs** = *cost of regenerated base oils sold – income from producers participating in the collective system.*

**On the second level of the model the 'social public cost'** is measured as:

**First, the possible lack of public effect (additional contribution) from the operation of most of the collective systems for utilization of waste oils.** According to the latest report published by the Executive Environment Agency (ExEA) in 2015, the percentage of recovered waste oils and waste oil products was 39% in 2009 and 44% in 2015 of the total quantity of these products released on the market in Bulgaria. There are two installations in the country with IPPC permits, both owned by commercial companies. One of them owns a collective recovery system as well. Their capacity exceeds the quantities reported in the EEA report (2015) as utilized by the total of 6 collective systems registered in Bulgaria and three companies performing individually their obligations. This leads to the conclusion that there is a high market demand for this quantity of base oil from producers of motor, transmission and industrial oils, lubricants, bitumen, and heavy fuel, the commercial companies with IPPC permits could collect and recover the waste oils on their own even if there are no collective systems. The role of the collective systems is only to obtain statements for the regenerated waste. According to information from the National Social Security Institute, at the end of 2017 five of the collective waste oil utilization systems registered in Bulgaria had 2 employees and only system had 4 employees. This proves that collective waste oil recovery systems lack the necessary capacity. This conclusion complements the problems of the corporate governance of the Bulgarian commercial companies described by Tchipev (2009) and Zahariev (2014) in a context different from the context of eco-business.

**Second, the inability of collective systems to raise sufficient funds in order to create their own technical capacity for recovery of waste.** If collective systems are allowed to create such capacity, the producers participating in these systems will pay lower fees and will have more money to

invest in modern resource- and energy-efficient production technologies. This would reduce the total amount of all harmful emissions and benefit the society. This conclusion could be supported by the hypothesis of Porter and Van der Linde (1995).

**Conclusion**

The number of research studies in the field of economic analysis of waste oil management is still insufficient. Moreover, the studies on environmental transaction costs focus mainly on policies implemented on a national level and the costs incurred by those affected by these policies. Our study focuses on the transaction costs of a certain participant in the implementation of the environmental policy and thus to complement the overall picture of the existing transaction costs in this field.

The lack of common and detailed European regulations regarding the collective systems bearing extended producer responsibility results in inefficient national legislations of the Member States and poses a risk of destroy the public trust in the waste management system.

The proposal for creation of a 'circular economy' (COM / 2015/0595), which is expected to enter into force after 2019, has partially addressed this issue.

Some of the positive aspects of the proposal are that it clarifies the ownership of collective systems and attempts to strengthen control over their financial management, but the provisions are too general and vague. Presently there are no specific audit standards for such control and one of the objectives of this study is to facilitate the development of such standards for the transaction costs of collective systems. We believe that most significant flaw of the new regulation is that it lacks incentives to encourage the collective systems to create their own technical capacity for recovery of waste. Although the collective systems accumulate a large amounts of financial resources, they have no right to share profits. The situation in Bulgaria shows that despite their large financial resources, they maintain a staff of 2 employees on average and the average value of their FTA is EUR 2000. At the same time, the almost equal utilization rate of waste oils for the last reported seven-year period (42% on average), as well as the capacity of the processing companies, which exceeds the utilized quantity, casts doubt on the contribution of the collective systems. The processing companies in Bulgaria carry out the full cycle of operations from collection of waste oils to production of finished products (base oils), which are no longer considered waste. Therefore, the collective systems do not participate with their own capacity at any stage of the entire recovery cycle and thus created conditions for generation of social public costs.

The methodology for reporting social costs in collective waste management systems is only marked as a basis for development of a second level of our model by using the output data from the transaction cost model for further analyses. Our final conclusion is that environmental transaction costs cannot be accounted separately from the related social costs and effects on public welfare.